\documentclass{article}
\usepackage{spconf,amsmath,graphicx}
\usepackage{amsfonts}
\usepackage{url}
\usepackage{tabularx}
\usepackage{siunitx}
\usepackage{multirow}
\usepackage{subfigmat}
\usepackage{mathtools}
\usepackage{color}

\renewcommand{\v}[1]{\ensuremath{\mathbf{#1}}}
\newcommand{\mat}[1]{\mathbf{#1}}

\newcommand{\argmax}{\mathop{\rm arg~max}\limits}

\title{End-to-end speaker diarization conditioned on\\speech activity and overlap detection}
\name{Yuki Takashima$^1$, Yusuke Fujita$^1$, Shinji Watanabe$^2$, Shota Horiguchi$^1$, Paola Garc\' ia$^2$, Kenji Nagamatsu$^1$}
\address{$^1$ Hitachi, Ltd. Research \& Development Group, Japan\\
$^2$ Center for Language and Speech Processing, Johns Hopkins University, USA}
\begin{document}
%
\maketitle
\begin{abstract}
In this paper, we present a conditional multitask learning method for end-to-end neural speaker diarization~(EEND).
The EEND system has shown promising performance compared with traditional clustering-based methods, especially in the case of overlapping speech.
In this paper, to further improve the performance of the EEND system, we propose a novel multitask learning framework that solves speaker diarization and a desired subtask while explicitly considering the task dependency.
We optimize speaker diarization conditioned on speech activity and overlap detection that are subtasks of speaker diarization, based on the probabilistic chain rule.
Experimental results show that our proposed method can leverage a subtask to effectively model speaker diarization, and outperforms conventional EEND systems in terms of diarization error rate.
\end{abstract}
\section{Introduction}
\label{sec:intro}
Speaker diarization is the process of partitioning a speech recording into homogeneous segments associated with each speaker.
This process is an essential part of multi-speaker audio applications such as generating transcriptions from meetings~\cite{Tranter2006,Anguera2012}.
Recent studies~\cite{conf/interspeech/KandaBHFHNH19,conf/asru/ZorilaBDH19} have shown that accurate diarization improves the performance of automatic speech recognition~(ASR).
Therefore, more advanced speaker diarization is required.

The traditional speaker diarization approach is a clustering-based method that relies on multiple steps: speech activity detection~(SAD), speech segmentation, feature extraction, and clustering.
SAD is the process to filter out the non-speech parts from an input speech.
Speech regions are then split into multiple speaker-homogeneous segments, and frame-level speaker embeddings are extracted.
The recent progress on deep learning has made it possible to compute better speaker representation such as x-vectors~\cite{conf/icassp/SnyderGSPK18} and d-vectors~\cite{conf/icassp/WanWPL18}.
Once, the embeddings are obtained, a clustering method is needed.
The methods commonly used are agglomerative hierarchical clustering~(AHC)~\cite{conf/slt/SellG14}, k-means clustering~\cite{conf/interspeech/DimitriadisF17}, spectral clustering~\cite{conf/icassp/WangDWMM18,conf/interspeech/LinYLBB19}, and affinity propagation~\cite{Frey07AffinityPropagation}.
Recently, neural network-based clustering has been explored~\cite{li2019discriminative}.
Although clustering-based methods performed well, they are not optimized to directly minimize diarization errors because the clustering is an unsupervised method.
To directly minimize diarization errors in a supervised manner, clustering-free methods have been studied~\cite{conf/icassp/ZhangWZP019,conf/interspeech/FujitaKHNW19,conf/asru/FujitaKHXNW19}.

End-to-end neural diarization~(EEND)~\cite{conf/interspeech/FujitaKHNW19,conf/asru/FujitaKHXNW19} is a promising direction for speaker diarization.
EEND uses a single neural network that maps a multi-speaker audio to joint speech activities of multiple speakers.
In contrast to most of the clustering-based methods, EEND handles overlapping speech without using any external module.
Most recently, EEND has been extended to handle a flexible number of speakers~\cite{journals/corr/abs-2006-01796,horiguchi2020end}.
Fujita {\it et al.}~\cite{journals/corr/abs-2006-01796} proposed speaker-wise conditional EEND~(SC-EEND) that produces a speech activity of one speaker conditioned by speech activities of previously estimated speakers.
However, the performance of SC-EEND is not enough for real recordings of conversation due to the low performance of SAD.
To tackle this problem, we employ a multitask learning approach that optimizes not only speaker diarization but its subtask, such as SAD and overlap detection~(OD).

In this paper, we propose a conditional multitask learning framework with the subtask of speaker diarization for SC-EEND.
Multitask learning~\cite{caruana_multitask_1997,zhang2018overview} is a learning strategy where a model simultaneously executes multiple tasks for one input data.
In this work, we utilize easier subtasks of speaker diarization such as SAD and OD as additional tasks.
It is expected that these subtasks have useful information to solve speaker diarization.
Moreover, there is an explicit dependency between speaker diarization and the subtasks.
Therefore, we propose the \textit{subtask-first speaker diarization model} that executes the subtask and speaker diarization successively,
similar to the coarse-to-fine approach on computer vision~\cite{Bar03}.

\begin{figure}[tb]

\begin{minipage}[b]{0.48\linewidth}
  \centering
  \centerline{\includegraphics[clip,width=3.8cm]{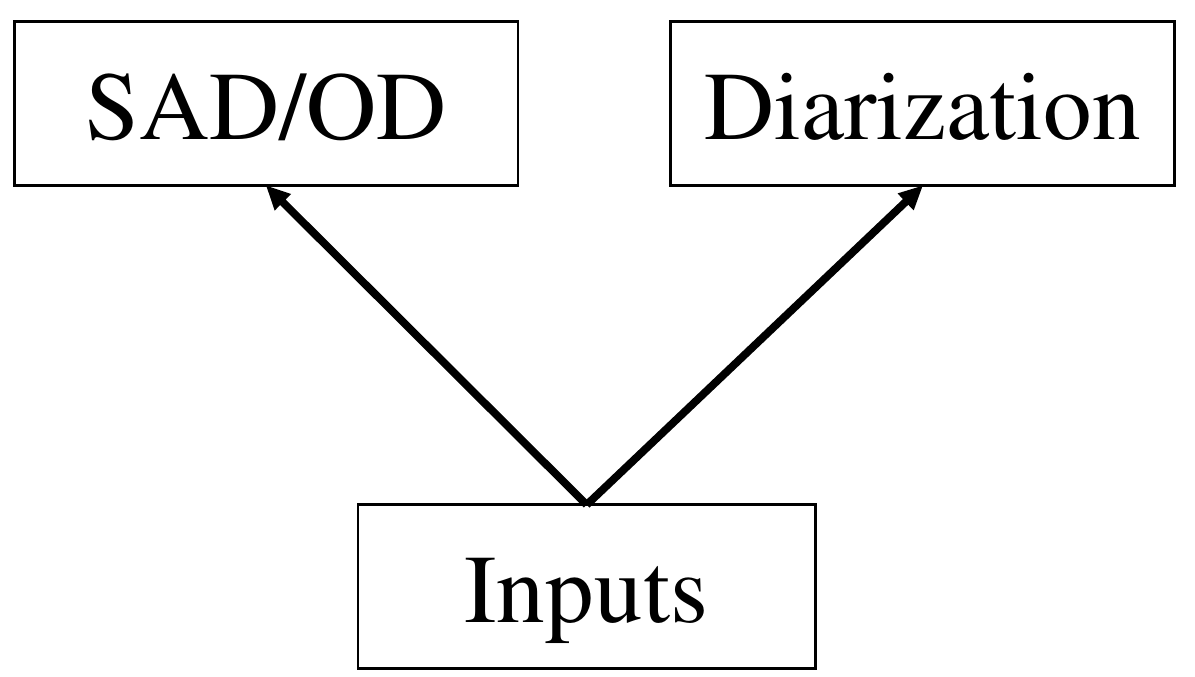}}
  \centerline{(a) Traditional approach}\medskip
\end{minipage}
\hfill
\begin{minipage}[b]{0.48\linewidth}
  \centering
  \centerline{\includegraphics[clip,width=3.8cm]{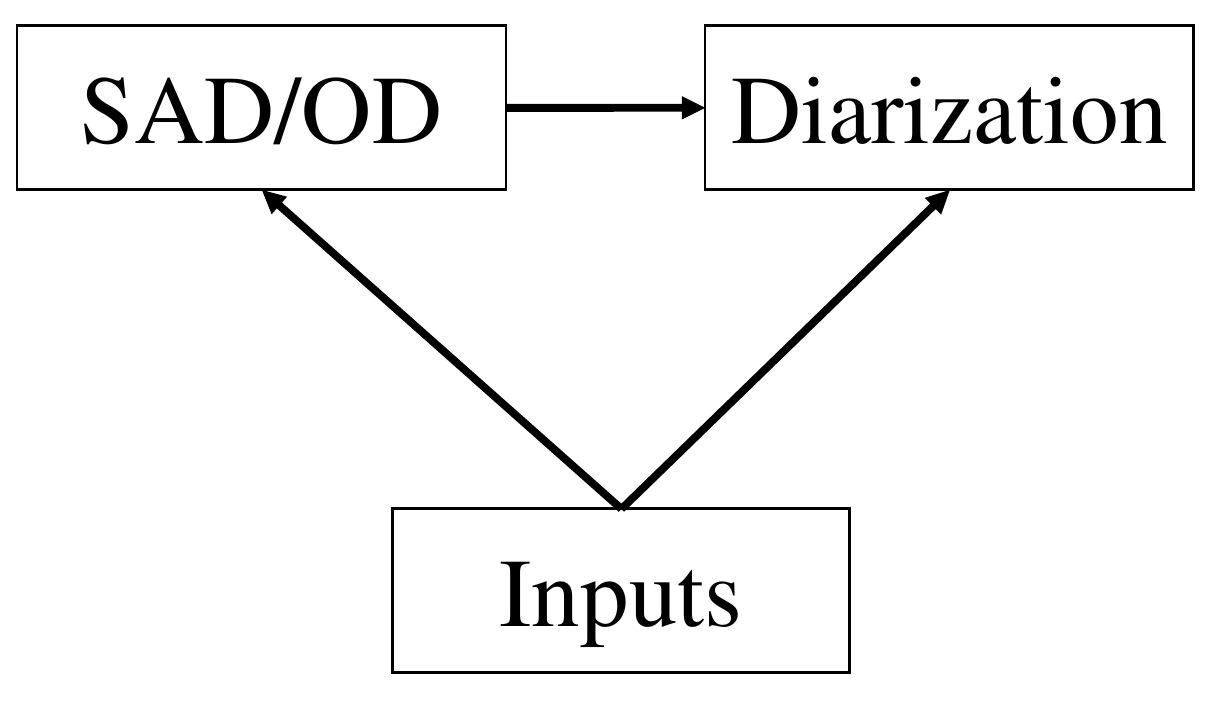}}
  \centerline{(b) Proposed approach}\medskip
\end{minipage}
\vspace{-5pt}
\caption{Multitask learning approaches.}
\label{fig:multitask}
\end{figure}

To take the order between tasks into account, we employ a conditional parallel mapping~\cite{shi2020sequence} that models the relevance between multiple output sequences explicitly via the probabilistic chain rule.
We model speaker diarization conditioned on a subtask via conditional parallel mapping as shown in Fig.~\ref{fig:multitask}, where both tasks are optimized simultaneously.
Our proposed method imitates the process of speaker diarization, which allows the model to learn effective dependency between tasks to improve the performance of speaker diarization.

In experiments, we perform three instances of our proposed approach: SAD-first SC-EEND, OD-first SC-EEND, and SAD-OD-first SC-EEND.
SAD-first SC-EEND and OD-first SC-EEND utilize SAD and OD as subtasks, respectively.
SAD-OD-first SC-EEND solves each task one by one in the order of SAD, OD, and speaker diarization, conditioned on the previous task.
It is an important property that multiple tasks can be combined.
The experimental results on CALLHOME~\cite{callhome} and simulated mixture datasets reveal that our proposed method achieves improvement over the conventional SC-EEND method.

The rest of this paper is organized as follows: In Section 2, related works is described.
In Section 3, our proposed method is described.
In Section 4, the experimental data are evaluated, and the final section is devoted to our conclusions.

\section{Related work}
\label{sec:related}

Speaker diarization has attracted attention because it can be used to boost the performance of ASR~\cite{conf/interspeech/KandaHTFNW19}.
Motivated by the CHiME Challenges~\cite{conf/interspeech/BarkerWVT18,watanabe2020chime} and the DIHARD Challenges~\cite{ryant2018first,ryant2019second}, several researchers have worked on developing more advanced speaker diarization system.
Lin~{\it et al.} proposed a long short-term memory~(LSTM)-based similarity measurement for the clustering-based speaker diarization.
Moreover, speech activity estimation based on neural networks has been proposed~\cite{ding2019personal,medennikov2020target} that directly produce the speech activity form the acoustic feature.
Kinoshita~{\it et al.} proposed all-neural model that jointly solves speaker diarization, source separation, and source counting and demonstrated the performance on real meeting scenarios.
On the other hand, the EEND~\cite{conf/asru/FujitaKHXNW19} has acquired significant interest as it can handle overlapping speech.
Various extensions of EEND had been proposed.
In~\cite{journals/corr/abs-2006-01796,horiguchi2020end}, the authors show experimental analyses and proposals for increasing the number of speakers.
In~\cite{xue2020online}, online speaker diarization based on EEND has been investigated using a speaker-tracing buffer.
In this paper, we focus on SC-EEND~\cite{journals/corr/abs-2006-01796} because it is highly expandable.

Under the scope of our research, multitask learning intends to leverage the useful information contained in multiple tasks to improve the generalization performance of those tasks.
Especially, there are some multitask learning approaches in which explicitly leverage the hierarchical relationship between tasks~\cite{conf/nips/ZhangWY18,journals/tip/0001ZKZZYP17}.
Sanh~{\it et al.}~\cite{conf/aaai/SanhWR19} proposed a hierarchical multitask learning model focused on semantic tasks, which describes information flow from the lower level tasks to more complex tasks in the single neural network.
Li~{\it et al.}~\cite{conf/icassp/LiKZ18} proposed a joint multitask learning framework that leverages the correlation between intent prediction and slot filling.
In the field of ASR, some recognized symbols have been used for multitask learning~\cite{conf/icassp/ChenMLS14,48982}.
Kubo~{\it et al.}~\cite{48982} proposed a joint phoneme-grapheme model that simultaneously converts a speech signal to phonemes and graphemes.
These methods consider an internal representation shared between tasks; however, the task-dependency cannot be considered.
In this paper, we investigate how to incorporate the relationship between the target task and its subtask into the model via the conditional chain mapping~\cite{shi2020sequence}.

\begin{figure*}[tb]
    \centering
    \subfigure[SC-EEND]{
        \includegraphics[clip,keepaspectratio, scale=0.34]{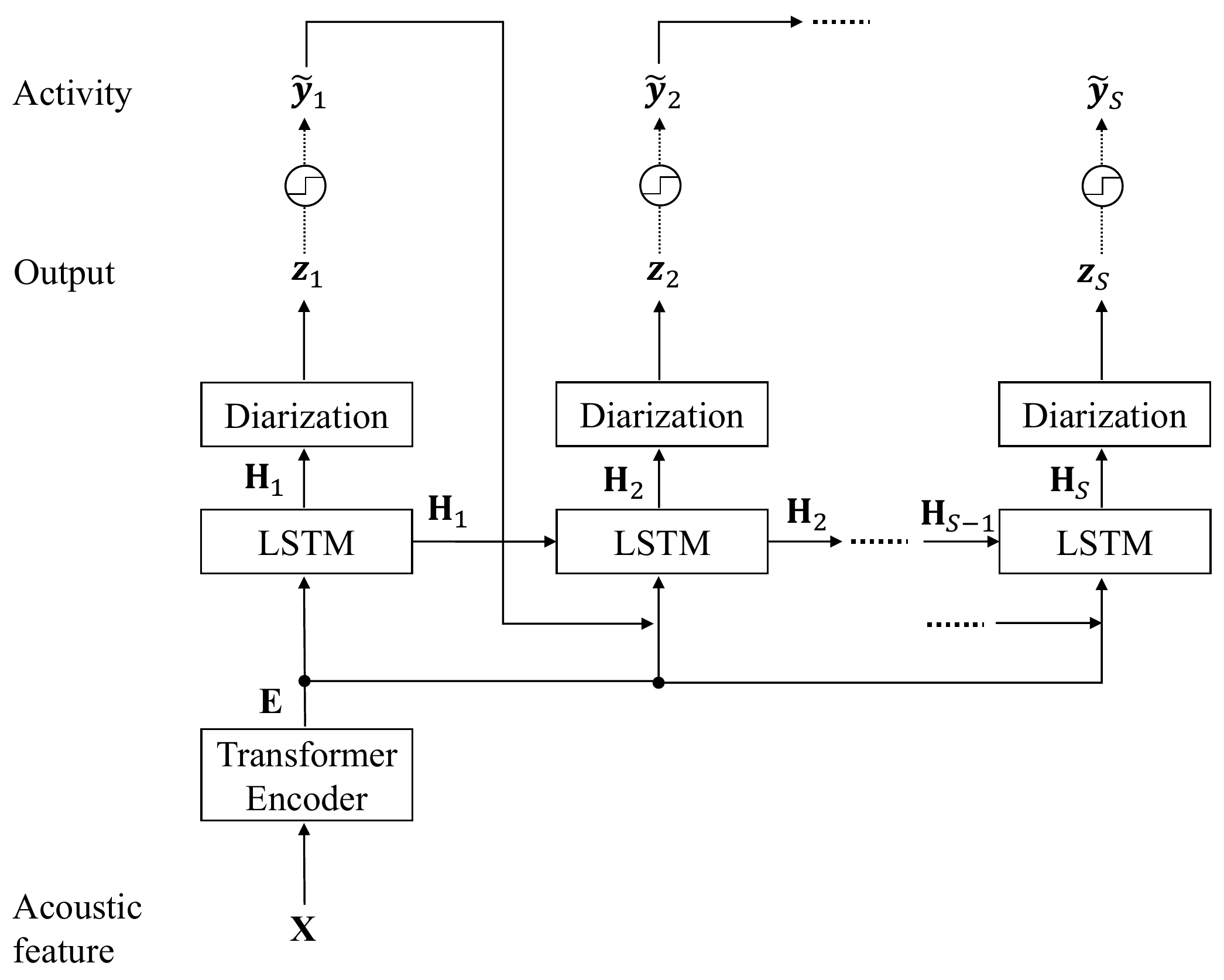}
        \label{fig:conventional_sceend}
    }
    \hfill
    \subfigure[Subtask-first SC-EEND]{
        \includegraphics[clip,keepaspectratio, scale=0.34]{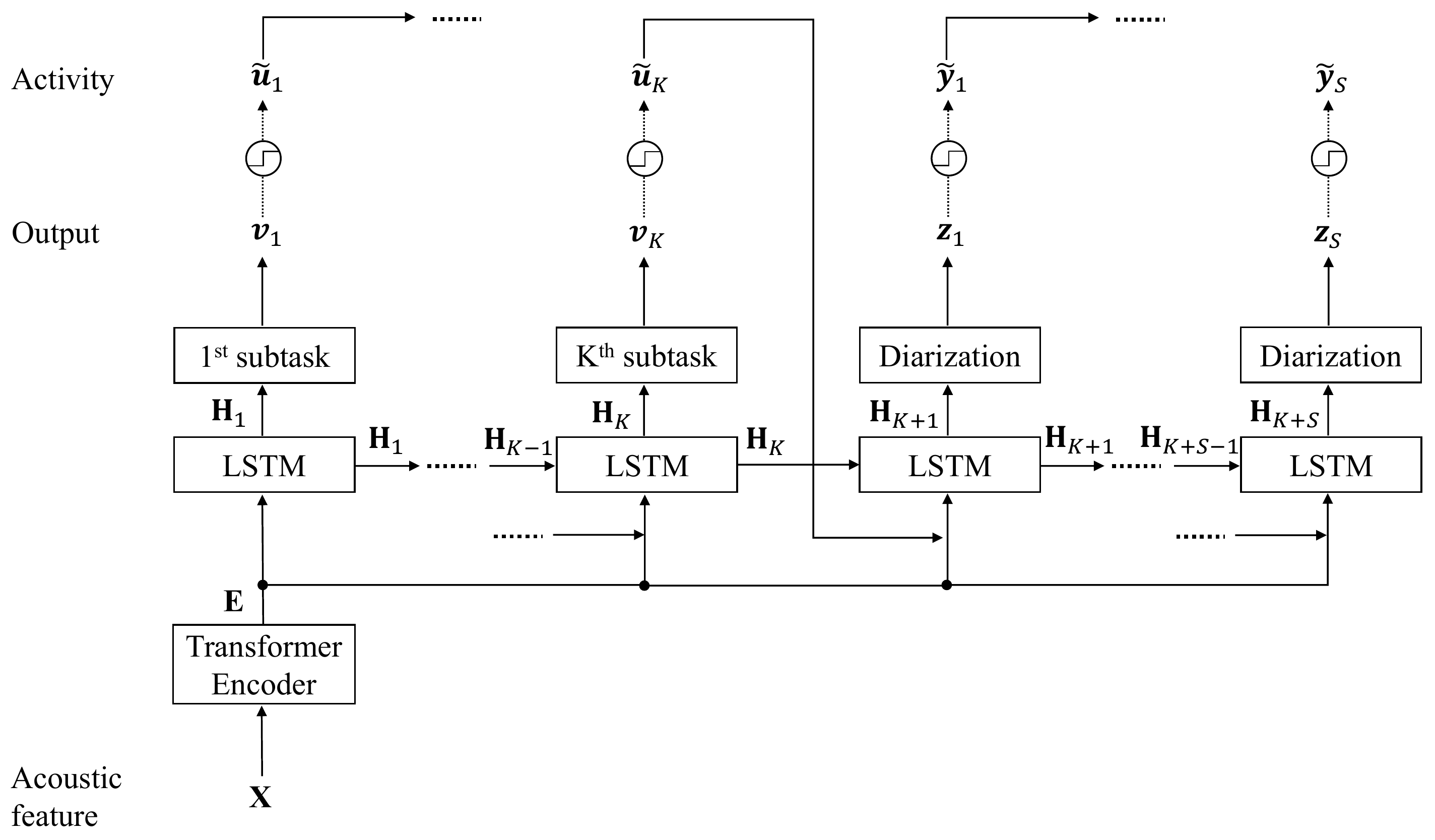}
        \label{fig:prop}
    }
    \vspace{-2pt}
    \caption{Overview of the conventional SC-EEND and proposed subtask-first SC-EEND.}
\end{figure*}

\section{Proposed method}
\label{sec:prop}

In this section, we describe subtask-first SC-EEND as an extension of SC-EEND~\cite{journals/corr/abs-2006-01796} shown in Fig.~\ref{fig:conventional_sceend}.
We formulate speaker diarization as a probabilistic model and then introduce the multitask mechanism using the probabilistic chain rule.
During training, our proposed method optimizes not only the diarization loss but the subtask loss by conditioning the speaker diarization on the subtask.
In this work, we employ SAD and OD as the subtasks.

\subsection{Subtask-first SC-EEND}
\label{sec:sad-sc-eend}
Given a $T$-length time sequence of $F$-dimensional acoustic features as a matrix $\mat{X}\in \mathbb{R}^{F\times T}$ and a set of speech activities $\left\{\v{y}_s\mid s\in \left\{1,\dots,S\right\}\right\}$ for speaker index $s$ and the number of speakers $S$, speaker diarization is formulated as follows:
\begin{equation}
\label{eq.diar}
\hat{\v{y}}_1, \dots, \hat{\v{y}}_S  = \argmax_{\v{y}_1, \dots, \v{y}_S} P(\v{y}_1, \dots, \v{y}_S\mid\mat{X}),
\end{equation}
where $\v{y}_s\in\{0,1\}^{1\times T}$ is a vector representing a time sequence of speech activity for speaker index $s$.
Eq.~\eqref{eq.diar} means that the most likely speaker activity results are estimated from the joint probability of $P(\v{y}_1, \dots, \v{y}_S\mid\mat{X})$.
In this paper, we extend Eq.~\eqref{eq.diar} by introducing $K$ subtasks and formulate a multitask problem of speaker diarization and the subtasks as
\begin{align}
&\hat{\v{y}}_1,\dots,\hat{\v{y}}_S,\hat{\v{u}}_1,\dots,\hat{\v{u}}_K \nonumber\\
=&\argmax_{\v{y}_1,\dots,\v{y}_S,\v{u}_1,\dots,\v{u}_K} P(\v{y}_1,\dots,\v{y}_S,\v{u}_1,\dots,\v{u}_K\mid\mat{X}),
\label{eq:multitask_formulation}
\end{align}
where $\v{u}_k=\{0,1\}^{1\times T}$ denotes an output for $k$-th subtask.
The joint probability in Eq. (\ref{eq:multitask_formulation}) can be transformed by using the probabilistic chain rule as follows:
\begin{align}
&P(\v{y}_1,\dots,\v{y}_S,\v{u}_1,\dots,\v{u}_K\mid\mat{X})\\
=&\prod_{k=1}^{K}\underbrace{P(\v{u}_k\mid \v{u}_1,\dots,\v{u}_{k-1},\mat{X})}_{\text{subtask distribution}}\nonumber\\
&\times\prod_{s=1}^{S}\underbrace{P(\v{y}_s\mid \v{y}_1,\dots,\v{y}_{s-1},\v{u}_1,\dots,\v{u}_K,\mat{X})}_{\text{speaker activity distribution}}\label{eq:prop_multitask}
\end{align}
Thus, we can factorize the joint probability by considering the dependency of the subtask and speaker activities.
The following explanations provide an actual form of subtask distribution $P(\v{u}_k\mid \v{u}_1,\dots,\v{u}_{k-1},\mat{X})$ and speaker activity distribution $P(\v{y}_s\mid \v{y}_1,\dots,\v{y}_{s-1},\v{u}_1,\dots,\v{u}_K,\mat{X})$, respectively.

Note that the order of $\v{y}_1,\dots,\v{y}_S,\v{u}_1,\dots,\v{u}_K$ is not unique, and it can be reordered.
This paper pre-define the part of the order (first subtask followed by speech activities) to realize subtask-first speaker diarization.

In the proposed method, we model both subtask and speaker activity probability distribution in Eq. (\ref{eq:prop_multitask}) as stateful neural network functions.
Each subtask and speaker activity distribution is represented as follows:
\begin{align}
\v{v}_k &= \mathrm{NN}_{\mathrm{Sub}}^{(k)}(\mat{X},\tilde{\v{u}}_{k-1}), \label{eq.output_sub}\\
\label{eq.output_diar}
\v{z}_s &= \mathrm{NN}_{\mathrm{Diar}}(\mat{X}, \tilde{\v{y}}_{s-1}),
\end{align}
where $\mathrm{NN}_{\mathrm{Sub}}^{(k)}(\cdot,\cdot)$ is a neural network that outputs a probability $\v{v}_k\in(0,1)^{1\times T}$ of the $k$-th subtask given an input $\mat{X}$ and the estimation of $(k-1)$-th subtask $\tilde{\v{u}}_{k-1}$.
$\mathrm{NN}_{\mathrm{Diar}}(\cdot,\cdot)$ is a speaker-wise conditional neural network that outputs a probability $\v{z}_s\in(0,1)^{1\times T}$ given $\mat{X}$ and the estimated $(s-1)$-th speaker's speech activity $\tilde{\v{y}}_{s-1}$.
To calculate the estimation of $k$-th subtask activity $\tilde{\v{u}}_k$ from the neural network output $\v{v}_k\coloneqq[v_{k,t}]_{t=1}^T$, we simply threshold each element in $\v{v}_k$ as
\begin{align}
    \tilde{\v{u}}_k=\left[\mathbb{I}(v_{k,t}>0.5)\right]_{t=1}^{T},
\end{align}
where $\mathbb{I}({\it cond})$ is an indicator function that takes $1$ if the condition ${\it cond}$ is true and $0$ otherwise.
The $s$-th speaker's speech activity $\tilde{\v{y}}_s$ is estimated from $\v{z}_s\coloneqq[z_{s,t}]_{t=1}^T$ in the same manner as
\begin{align}
    \tilde{\v{y}}_s=\left[\mathbb{I}(z_{s,t}>0.5)\right]_{t=1}^{T}.
\end{align}
Note that the condition for the first subtask network $\tilde{\v{u}}_0$ is a zero matrix $\mat{0}^{(1\times T)}$ and that used to extract the first speaker is the estimation of the last, i.e. $K$-th, subtask: $\tilde{\v{y}}_{0}\coloneqq\tilde{\v{u}}_K$.

In this paper, we adopt SAD and OD as subtasks. The reference labels for SAD and OD are uniquely determined by those of diarization $\v{y}^{*}_s\coloneqq[y^{*}_{s,t}]_{t=1}^{T}\in\{0,1\}^{1\times T}$ as follows:
\begin{align}
    \v{u}^{*}_k\coloneqq\begin{cases}
        \left[\max(y^{*}_{1,t}, \dots, y^{*}_{S,t})\right]_{t=1}^{T}&(\text{if $k$-th subtask is SAD})\\
        \left[\mathbb{I}\left(\sum_{s=1}^{S} y^{*}_{s,t}>1\right)\right]_{t=1}^{T}&(\text{if $k$-th subtask is OD})
    \end{cases}.
    \label{eq:subtask_ref_label}
\end{align}
Therefore, we do not need additional labeling to carry out the proposed multitask learning approach.
The detailed implementation of the proposed model is described in the following section.

\subsection{Model design}

Figure~\ref{fig:prop} shows an overview of the proposed method.
In order to model the task relationship, we employ a conditional chain mapping~\cite{shi2020sequence} that explicitly models the relevance between multiple output sequences with the probabilistic chain rule.
In~\cite{shi2020sequence}, its effectiveness has been proven on speech separation and multi-speaker speech recognition where each of the output sequences corresponds to the speech or the text of each individual speaker.
In this work, we regard the SAD/OD activities and speech activities of each speaker as an additional sequence along the subtask $k$ and speaker $s$ direction.
Our model consists of three parts: shared encoder, conditional chain module, and task-specific decoder.

For the encoder part, similar to SC-EEND \cite{journals/corr/abs-2006-01796}, we use four stacked Transformer encoder blocks \cite{Vaswani2017}, each of which consists of a multi-head self-attention layer, a position-wise feed-forward layer, and residual connections.
In this step, the input sequence of acoustic features $\mat{X}\in \mathbb{R}^{F\times T}$ is mapped to a sequence of $D$-dimensional embeddings $\mat{E}\in \mathbb{R}^{D\times T}$.
This encoder is shared among $\mathrm{NN}_\mathrm{Sub}^{(k)}(\cdot,\cdot)$ and $\mathrm{NN}_\mathrm{Diar}(\cdot,\cdot)$ in Eqs. (\ref{eq.output_sub}) and (\ref{eq.output_diar}).

For the conditional chain module, similar to \cite{shi2020sequence}, we use a uni-directional LSTM, whose hidden representation at $l$-th step $\mat{H}_l\in\mathbb{R}^{D\times T}$ is calculated one by one as
\begin{align}
&\mat{H}_l=\nonumber\\
&\begin{cases}
\mathrm{LSTM}([\begin{matrix}\mat{E},f(\tilde{\v{u}}_{l-1})\end{matrix}], \mat{H}_{l-1})&(1\leq l\leq K)\\
\mathrm{LSTM}([\begin{matrix}\mat{E},f(\tilde{\v{y}}_{l-K-1})\end{matrix}], \mat{H}_{l-1})&(K+1\leq l\leq K+S)\\
\end{cases},
\end{align}
where $f(\cdot)$ is a linear projection that maps a scalar to a $D$-dimensional vector for each row of the input vector.
$\mathrm{LSTM}(\cdot,\cdot)$ maps a $2D$-dimensional vector to a $D$-dimensional vector while keeping a $D$-dimensional memory cell for each column of the input matrix.
The initial hidden representation $\mat{H}_0$ is a zero matrix $\mat{0}^{(D\times T)}$.

For the decoder part, the output probabilities $\v{v}_k$ and $\v{z}_s$ in Eqs. (\ref{eq.output_sub}) and (\ref{eq.output_diar}) are computed as follows:
\begin{align}
\label{eq:fc_sub}
\v{v}_k &= \sigma(g^{(k)}_\mathrm{Sub}(\mat{H}_k)), \\
\v{z}_s &= \sigma(g_\mathrm{Diar}(\mat{H}_{s+K})),
\end{align}
where $g^{(k)}_{\mathrm{Sub}}(\cdot)$ and $g_{\mathrm{Diar}}(\cdot)$ are linear projections for the $k$-th subtask and speaker diarization, respectively.
These functions map a $D$-dimensional vector to a scalar for each column of the input matrix.
$\sigma(\cdot)$ is a sigmoid activation function that produces an output probability.

Our proposed model is optimized minimizing the losses for the subtask and speaker diarization.
The subtask loss between the neural network output $\v{v}_k$ and reference label $\v{u}_k^{*}$ determined in Eq. (\ref{eq:subtask_ref_label}) is computed as follows:
\begin{align}
L_{\mathrm{Sub}} = \frac{1}{T}\sum_{k=1}^{K}\mathrm{BCE}(\v{v}_k, \v{u}_k^{*}),
\end{align}
where $\mathrm{BCE}(\cdot,\cdot)$ is an element-wise binary cross-entropy function followed by the summation of all elements.
For speaker diarization, the loss between the neural network output $\v{z}_s$ and reference label $\v{y}_s^{*}$ is calculated as
\begin{align}
\label{eq.pit}
L_{\mathrm{PIT}} = \frac{1}{ST}\min_{\phi \in \mathrm{perm}(S)}\sum_{s=1}^S \mathrm{BCE}(\v{z}_s, \v{y}^{*}_{\phi_s}),
\end{align}
where $\mathrm{perm}(S)$ is a set of all possible permutations of a sequence $(1, \dots, S)$, and $\v{y}^{*}_{\phi_s}$ indicates the $s$-th reference label after the permutation $\phi$.
This loss is referred as permutation-invariant training~(PIT) loss~\cite{conf/icassp/YuKT017}.
Finally, the loss of our proposed model is written as:
$L = L_{\mathrm{Sub}} + L_{\mathrm{PIT}}.$
During training, we use the teacher-forcing~(TF) technique~\cite{WilliamsZipser:89nc} to boost the performance by exploiting the ground-truth labels, i.e., $\tilde{\v{u}}_{k-1}$ and $\tilde{\v{y}}_{s-1}$ in Eqs. (\ref{eq.output_sub}) and (\ref{eq.output_diar}) are replaced by $\v{u}^{*}_{k-1}$ and $\v{y}^{*}_{s-1}$, respectively.
However, the order of the speakers cannot be determined in advance because it is determined during training. To alleviate this problem, we examine the two-stage PIT loss computation strategy~\cite{journals/corr/abs-2006-01796}.
This technique firstly estimates speech activities of all speakers.
Then, we obtain an optimal permutation by calculating the PIT loss using Eq. (\ref{eq.pit}).
In the next step, network outputs are produced again using ground-truth labels associated with the optimal permutation.
We also use a stop sequence condition to handle variable number of speakers as in the original SC-EEND paper \cite{journals/corr/abs-2006-01796}.

\section{Experimental results}
\label{sec:exp}

\subsection{Conditions}
\label{sec:cond}

\begin{table}[tb]
\vspace{-8pt}
\caption{Statistics of training/adaptation/test sets.}
\vspace{5pt}
\label{tbl:set}
\centering
\scalebox{0.9}{
\begin{tabular}{@{}lrrrr@{}} \hline\hline
& Num. & Num. of & Avg. dur. & \multicolumn{1}{c}{Overlap} \\
& spk & mixtures & (sec) & ratio (\%) \\\hline
Traning sets & & & & \\
\:Simulated-2spk & 2 & 100,000 & 88.6 & 34.1 \\
\:Simulated-vspk & 1-4 & 100,000 & 130.0 & 29.7 \\
\hline
Adaptation sets & & & & \\
\:CALLHOME-2spk & 2 & 155 & 74.0 & 14.0 \\
\:CALLHOME-vspk & 2-7 & 249 & 125.8 & 17.0 \\
\hline
Test sets & & & & \\
\:CALLHOME-2spk & 2 & 148 & 72.1 & 13.0 \\
\:CALLHOME-vspk & 2-6 & 250 & 123.2 & 16.7 \\ \hline\hline
\end{tabular}
}
\vspace{-4pt}
\end{table}

The proposed method was evaluated for both two-speaker and variable-speaker audio mixtures.
We prepared a simulated training set based on~\cite{conf/asru/FujitaKHXNW19}.
We also prepared real adaptation/test sets from CALLHOME~\cite{callhome}.
The statistics of the datasets are listed in Table~\ref{tbl:set}.
For the CALLHOME-2spk and CALLHOME-vspk sets, we employed identical sets as the ones given in Kaldi CALLHOME\_diarization v2 recipe\footnote{https://github.com/kaldi-asr/kaldi/tree/master/egs/callhome\_diarization},
to ensure a fair comparison with the x-vector clustering method~\cite{Sell2018dihard}.
The recipe uses AHC with probabilistic linear discriminant analysis~(PLDA) scoring scheme.
In this case, time-delay neural network-based speech activity detection~(TDNNSAD) was used.
The number of clusters was fixed to two for the two-speaker experiments.
On the other hand, to estimate the variable number of speakers the PLDA scores were used.

We also compared the proposed method with two conventional EEND-based systems: SA-EEND system and SC-EEND system.
Additionally, we also evaluated the traditional multitask learning method to verify the effectiveness of the proposed conditional connections.
A multitask model~(Multitask SC-EEND) has a task-specific LSTM without the conditional connection between two tasks, unlike the subtask-first SC-EEND model as shown in Fig.~\ref{fig:multitask} (a).
Moreover, we investigated an application of our proposed method on a three-task scenario where SAD and OD and speaker diarization are solved in this order.
For the EEND-based systems, including the proposed system,
the input features were 23-dimensional log-Mel-filterbanks with a \SI{25}{\ms} frame length and \SI{10}{\ms} frame shift.
For the two-speaker experiments, each feature was concatenated with those from the previous seven frames and subsequent seven frames.
After subsampling the concatenated features by a factor of ten.
We used four encoder blocks with 256 attention units containing four heads.
For variable number of speaker experiments, we used a concatenation length of 29 and a subsampling ratio of 20 which are twice larger than that of two-speaker experiments.
We used four encoder blocks and with 384 attention units containing six heads.

We used diarization error rate~(DER) as evaluation metric.
A \SI{250}{\ms} collar was employed at the start and end of each segment. Note that we included errors in overlapped segments and SAD-related errors for the DER calculation, whereas most works in literature did not evaluate such errors.

\subsection{Results and discussion}

\subsubsection{Experiments on fixed two-speaker models}

\begin{table}[t]
\vspace{-8pt}
\caption{Detailed DERs (\%) evaluated on CALLHOME-2spk. DER is composed of Misses (MI), False alarms (FA), and Confusion errors (CF). The SD errors are composed of Misses (MI) and False alarms (FA) errors.
}
\vspace{3pt}
\label{tbl:2spk}
\centering
\setlength{\tabcolsep}{4.4pt}
\begin{tabular}{@{}l|c|ccc|cc@{}} \hline\hline
& & \multicolumn{3}{c|}{DER breakdown} & \multicolumn{2}{c}{SAD} \\
Method & DER & MI & FA & CF & MI & FA \\ \hline
Clustering-based &&&&& \\
\: i-vector & 12.10 & 7.74 & 0.54 & 3.82 & 1.4 & 0.5 \\
\: x-vector & 11.53 & 7.74 & 0.54 & 3.25 & 1.4 & 0.5 \\ \hline
EEND-based & &&&&&\\
\: SA-EEND & 10.32 & 5.66 & 3.25 & 1.40 & 3.0 & 0.5  \\
\: SC-EEND & 9.39 & 4.96 & 2.73 & 1.70 & 2.2 & 0.4 \\ \hline
SC-EEND w/ subtask & &&&&&\\
\: SAD-multitask & 9.20 & 4.89 & 2.61 & 1.70 & 2.2 & 0.4 \\
\: SAD-first & 8.81 & 4.11 & 2.96 & 1.74 & 1.4 & 0.8 \\
\: OD-multitask & 9.16 & 4.59 & 2.52 & 2.05 & 2.2 & 0.5 \\
\: OD-first & 9.09 & 5.25 & 1.86 & 1.98 & 2.3 & 0.4  \\
\: SAD-OD-multitask & 9.06 & 4.90 & 2.35 & 1.80 & 2.3 & 0.5 \\
\: SAD-OD-first & {\bf 8.53} & 4.22 & 2.33 & 1.98 & 1.6 & 0.7  \\
\hline\hline
	\end{tabular}
	\vspace{-9pt}
\end{table}

First, we confirmed the behavior of the model trained on two-speaker mixtures as shown in Table~\ref{tbl:2spk}.
As observed, DERs on EEND-based systems outperformed the conventional clustering-based methods.
The main reason is that EEND can handle overlapping speech.

Second, we observed the performance of the models using SAD as the subtask.
SAD-first SC-EEND achieved a 6.2\% relative improvement over SC-EEND.
Compared with the traditional multitask model, we observed a 4.2\% relative improvement.
Another important observation is that our proposed method significantly reduced the MI errors in SAD errors compared with traditional multitask learning.
Surprisingly, our models achieved comparable performance with clustering-based methods that have a SAD module trained separately.
This result indicates that our proposed method can improve the subtask's performance and can leverage it to the subsequent task effectively via the conditional mechanism.

Next, we discuss the performance of models using OD as the subtask.
The OD-first SC-EEND outperformed both SC-EEND and traditional multitask models, and showed slightly worse performance than SAD-first SC-EEND.
On the other hand, the FA errors in DER breakdown were significantly reduced.
Considering the result, we hypothesized that the overlap information helped the model prevent overproducing the overlap speech activity in single-speaker segments.

Furthermore, we describe the results on the multiple subtask scenario with SAD and OD.
As shown in the table, SAD-OD-first SC-EEND achieved the lowest DER among the evaluated methods.
By utilizing two subtasks, SAD-OD-first SC-EEND showed 3.18\% and 6.16\% relative DER improvements over SAD-first and OD-first approaches, respectively.
Similar to SAD-first SC-EEND, we observed a significant reduction in SAD errors.
The results also showed that FA errors in DER breakdown were significantly reduced compared with SAD-first SC-EEND owing to the conditioning on OD.
These results indicate that conditioning speaker diarization on the subtasks contributes the significant performance improvement.
Note that an increase of the number of parameters is less than 0.01\% for the SAD-OD-first model associated with Eq.~(\ref{eq:fc_sub}) whereas about 17\% for the multitask model.
This means that our proposed method is an efficient strategy to model the relationship between tasks.

\subsubsection{Experiments on variable number of speakers}

\begin{table}[t]
\vspace{-8pt}
\caption{DERs (\%) on CALLHOME-vspk.}
\vspace{3pt}
\label{tbl:vspk}
\centering
\begin{tabular}{@{}lc@{}} \hline\hline
Method &  DER \\ \hline
Clustering-based & \\
\: x-vector & 19.01  \\ \hline
EEND-based & \\
\: SC-EEND & 15.57 \\
\: SAD-first SC-EEND & 15.36 \\
\: OD-first SC-EEND & 16.37 \\
\: SAD-OD-first SC-EEND & \bf 15.32\\
\hline\hline
	\end{tabular}
	\vspace{-9pt}
\end{table}

DERs on variable number of speakers for CALLHOME test set are shown in Table~\ref{tbl:vspk}.
For the adaptation of SAD-first and SAD-OD-first models, we dropped the SAD subtask losses at randomly selected frames with a ratio of 0.7 and multiplied the losses by 0.1, to prevent overfitting.
For the inference using the SAD-first model, we used the outputs of the SAD subtask network to determine non-speech frames regardless of the diarization outputs.
According to the table, SAD-first SC-EEND achieved 15.36\%, which corresponds to 19.2\% and 1.35\% relative DER improvements over the conventional x-vector clustering method and SC-EEND.
Moreover, SAD-OD-first SC-EEND reached 15.32\% DER which is the best performance among the evaluated methods.
These results indicate that our proposed SAD-first approach is also effective in a variable-speaker setting.
However, OD-first SC-EEND did not outperform the conventional SC-EEND although it outperformed the conventional x-vector clustering method.
It is assumed that a more careful training strategy is needed such as the scheduled learning because OD is a harder task than SAD.
We will investigate a more elaborated strategy to tacke this problem in the future.

Table~\ref{tab:detailspk} shows the detailed DER results of the best-proposed system result (SAD-OD-first SC-EEND) and the conventional SC-EEND in CALLHOME-vspk (Table~\ref{tbl:vspk}) for each number of speakers.
SAD-OD-first SC-EEND is better than the conventional SC-EEND in most cases except for the four-speaker case.
This result indicates that our proposed approach is robust to the increase in the number of speakers.

We also analyze the accuracy of speaker counting.
The results are shown in Table~\ref{tab:numberofspeaker}.
The proposed method achieved better speaker counting accuracy than the x-vector+AHC method; however, it was still difficult to handle more than four speakers.
One of the reasons is that the CALLHOME dataset consists of an imbalance number of speakers.
Therefore, it is needed to solve this problem using several training strategies such as generating the simulated data and applying the loss weighting.
Moreover, the conventional SC-EEND achieved slightly higher speaker counting accuracy than SAD-OD-first SC-EEND.
This indicates that SAD-OD-first SC-EEND focuses to solve diarization accurately for the small number of speakers, as shown in Table~\ref{tab:detailspk}.

Finally, we evaluated our proposed method with comparison to other systems as shown in Table~\ref{tbl:sota}. In this comparison, we only evaluated single-speaker regions.
For this purpose, we used oracle SAD and OD information as the conditions, and filtered out non-speech frames of the estimated diarization result using the oracle SAD information.
Although our proposed method could not achieve state-of-the-art performance, it outperformed the system of Zhang {\it et al.}~\cite{conf/icassp/ZhangWZP019}.
Compared to McCree's system~\cite{conf/interspeech/McCreeSG19}, our proposed method has an advantage that the system can be constructed as a single neural network without a complex implementation.
Our proposed method achieved competitive performance with other systems which also suggests that it can use external subtask information via the subtask-first model.

\begin{table}[tb]
\vspace{-8pt}
    \caption{Detailed DERs (\%) associated with each number of speaker on CALLHOME-vspk.}
    \vspace{3pt}
    \centering
    \setlength{\tabcolsep}{3.0pt}
    \begin{tabular}{l|ccccc} \hline\hline
    & \multicolumn{5}{c}{Num. of speakers} \\
    Model & 2 & 3 & 4 & 5 & 6 \\ \hline
    SC-EEND & 9.0 & 14.4 & \textbf{19.1} & 34.6 & 39.5 \\
    SAD-OD-first SC-EEND & \textbf{8.0} & \textbf{13.5} & 23.1 & \textbf{30.0} & \textbf{35.2} \\ \hline\hline
    \end{tabular}
    \label{tab:detailspk}
\end{table}

\begin{table}[tb]
\vspace{-10pt}
    \caption{Speaker counting results on variable-speaker CALLHOME. SAD-OD-first SC-EEND-based and SC-EEND-based models were trained with PIT+TF.}
    \vspace{5pt}
    \centering
    \begin{subfigmatrix}{2}
    \setlength{\tabcolsep}{5.1pt}
    \subtable[x-vector+AHC (Acc: 54.6\%)]{
    \begin{tabular}{cc|ccccc} \hline
     & & \multicolumn{5}{c}{Estimated} \\
     & & 2 & 3 & 4 & 5 & 6  \\ \hline
     \multirow{5}{*}{\rotatebox{90}{Reference}}
     &2 & \bf 84 & 62 & 2 & 0 & 0  \\
     &3 & 18 & \bf 51 & 5 & 0 & 0  \\
     &4 &  2 & 12 & \bf 6 & 0 & 0  \\
     &5 &  0 &  4 & 1 & \bf 0 & 0  \\
     &6 &  0 &  1 & 2 & 0 & \bf 0  \\ \hline
    \end{tabular}
    }
    \subtable[Proposed (Acc: 75.6\%)]{
    \begin{tabular}{c|ccccc} \hline
      & \multicolumn{5}{c}{Estimated} \\
      & 2 & 3 & 4 & 5 & 6  \\ \hline
     2 & \bf 129 & 19 & 0 & 0 & 0  \\
     3 & 15 & \bf 54 & 5 & 0 & 0 \\
     4 &  2 & 12 & \bf 6 & 0 & 0 \\
     5 &  0 & 2 & 3 & \bf 0 & 0  \\
     6 &  0 & 1 & 2 & 0 & \bf 0 \\ \hline
    \end{tabular}
    }
    \subtable[SC-EEND (Acc: 77.6\%)]{
    \begin{tabular}{c|ccccc} \hline
      & \multicolumn{5}{c}{Estimated} \\
      & 2 & 3 & 4 & 5 & 6  \\ \hline
     2 & \bf 129 & 19 & 0 & 0 & 0  \\
     3 & 9 & \bf 60 & 5 & 0 & 0 \\
     4 &  1 & 14 & \bf 5 & 0 & 0 \\
     5 &  0 & 1 & 4 & \bf 0 & 0  \\
     6 &  0 & 1 & 2 & 0 & \bf 0 \\ \hline
    \end{tabular}
    }
    \end{subfigmatrix}
    \label{tab:numberofspeaker}
    \vspace{-10pt}
\end{table}

\begin{table}[tb]
\vspace{-11pt}
\caption{DERs (\%) on CALLHOME-vspk with oracle SAD. Overlap segments were excluded from the DER calculation.
Note that the evaluation set used in the proposed method was different from other systems.
We used a random subset of CALLHOME, while other systems used the whole CALLHOME evaluation set.
}
\vspace{5pt}
\label{tbl:sota}
\centering
\begin{tabular}{lc} \hline\hline
Method & DER \\ \hline
McCree {\it et al.}~\cite{conf/interspeech/McCreeSG19} & 7.1 \\
SAD-OD-first SC-EEND & 7.4 \\
Zhang {\it et al.}~\cite{conf/icassp/ZhangWZP019} & 7.6 \\
\hline\hline
\end{tabular}
\vspace{-10pt}
\end{table}

\section{Conclusion}
\label{sec:conc}
In this paper, we proposed an end-to-end speaker diarization conditioned on specific subtasks.
Our proposed model performs not only speaker diarization but also SAD/OD based on the probabilistic chain rule.
In our experiments, we confirmed that our proposed subtask-first SC-EEND improves the DER on the two-speaker CALLHOME dataset compared with EEND-based systems with the traditional multitask learning.
Furthermore, our proposed SAD-first approaches showed the robustness to increasing the number of speakers.

In the future, we will investigate the combination with other tasks.
Moreover, we will explore the variable number of speaker and other difficult scenarios.

\vfill

\bibliographystyle{IEEEbib}
\bibliography{refs}

\end{document}